\documentclass{article}
\usepackage{ijcai26}

\usepackage{times}
\usepackage{soul}
\usepackage{url}
\usepackage[hidelinks]{hyperref}
\usepackage[utf8]{inputenc}
\usepackage[small]{caption}
\usepackage{graphicx}
\usepackage{amsmath}
\usepackage{amsfonts}
\usepackage{amsthm}
\usepackage{booktabs}
\usepackage{multirow}
\usepackage{algorithm}
\usepackage{algorithmic}
\usepackage[switch]{lineno}

\title{Repurposing Backdoors for Good: Ephemeral Intrinsic Proofs for Verifiable Aggregation in Cross-silo Federated Learning}

\author{
Xian Qin$^1$
\and
Xue Yang$^{1*}$\and
Xiaohu Tang$^1$\And
\affiliations
$^1$Southwest Jiaotong University\\
\emails
xq@my.swjtu.edu.cn,
xueyang@swjtu.edu.cn,
xhutang@swjtu.edu.cn
}

\begin{document}

\maketitle

\begin{abstract}
While Secure Aggregation (SA) protects update confidentiality in Cross-silo Federated Learning, it fails to guarantee aggregation integrity, allowing malicious servers to silently omit or tamper with updates. Existing verifiable aggregation schemes rely on heavyweight cryptography (e.g., ZKPs, HE), incurring computational costs that scale poorly with model size.
In this paper, we propose a lightweight architecture that shifts from extrinsic cryptographic proofs to \textit{Intrinsic Proofs}. We repurpose backdoor injection to embed verification signals directly into model parameters. By harnessing Catastrophic Forgetting, these signals are robust for immediate verification yet ephemeral, naturally decaying to preserve final model utility.
We design a randomized, single-verifier auditing framework compatible with SA, ensuring client anonymity and preventing signal collision without trusted third parties.
Experiments on SVHN, CIFAR-10, and CIFAR-100 demonstrate high detection probabilities against malicious servers. Notably, our approach achieves over $1000\times$ speedup on ResNet-18 compared to cryptographic baselines, effectively scaling to large models.
\end{abstract}

\section{Introduction} 
\label{sec:intro}

Federated Learning (FL)~\cite{FedAvg} enables distributed participants to collaboratively train a model by exchanging model updates rather than raw data. 
While this offers a level of confidentiality, the aggregation process of these updates is unsupervised, as clients lack a mechanism to verify the correctness of the server's computation.
This vulnerability is particularly critical in cross-silo scenarios, where participants are distinct, mutually distrustful institutions (e.g., banks or hospitals). Cross-silo architectures frequently rely on an outsourced, third-party server to coordinate aggregation. This server acts merely as a coordinator rather than the model owner. Lacking a long-term stake in the global model's utility, such outsourced servers may be economically motivated to act maliciously, selectively omitting updates to reduce computational overhead or by sabotaging updates to favor specific institutional rivals. Such integrity breaches degrade model utility without detection \cite{VerifyNet,VeriFL,survey}. To address this trust deficit, clients require a mechanism to verify the honest inclusion of their local updates.

Existing verifiable aggregation rely on \textit{extrinsic cryptographic proofs}. These approaches treat verification as an external dependency distinct from the learning task. They employ heavyweight cryptographic primitives (e.g., Homomorphic Encryption, Zero-Knowledge Proofs (ZKPs), Cryptographic commitments) to construct proofs of inclusion, result in \textit{clients must generate and transmit a separate proof alongside local updates.} To complete verification, clients must execute complex algorithms to confirm that the aggregated proof aligns with the global model parameters~\cite{YANG2024238,Chen2025,VerifyNet}.
Despite their theoretical soundness, these approaches face three critical limitations:
(i) \textbf{Prohibitive Efficiency Overhead}: Generating and transmitting proofs proportional to model dimensionality incurs huge computational and communication burdens, causing existing schemes impractical for large-scale networks; and
(ii) \textbf{Restrictive Assumptions}: Many schemes require auxiliary verifiers or non-colluding multi-server setups.
\textit{These constraints highlight a need for a verification mechanism that is lightweight, scalable and independent of trusted third parties.}

To address these limitations, we propose a paradigm shift from heavy extrinsic cryptographic proofs to a lightweight Intrinsic Auditing Architecture. Our core insight is that the model parameters themselves can serve as the verification medium.
We replace external commitments with Intrinsic Proofs, which are verification signals injected directly into the local model parameters, rather than generated alongside the update. To realize this, we repurpose the mechanics of backdoor injection, transforming it from a persistent malicious attack into a constructive verification mechanism. Functionally, the backdoor serves as a specific input-output pattern; if a local model containing this pattern is honestly aggregated, the global model will exhibit a corresponding detectable response that reflects the same input-output pattern. Otherwise, the absence of  this response indicates omission. 
This detectable response serves as empirical evidence of inclusion, eliminating the overhead of separate proof transmission. Importantly, this Intrinsic Proof concept is fully compatible with Secure Aggregation (SA) protocols \cite{google2017,QIN2026}, which strengthen the privacy of clients by protecting local updates during aggregation.

However, existing backdoor mechanisms are primarily engineered for malicious attacks emphasizing persistence~\cite{Neurotoxin2022,PerDoor2023} or post-training, long-term ownership verification by a single owner~\cite{WAFFLE2021,SMC2021}. These persistence requirements make them ill-suited for the dynamic, iterative verification required in our context. In contrast, our framework necessitates an inverted design philosophy. To be effective, the Intrinsic Proof mechanism must satisfy two rigorous properties:
\begin{enumerate}
    \item Unlike ownership verification that demand permanence, the Intrinsic Proof require ephemeral. It requires robust detectability immediately after the aggregation, yet must decay during subsequent training. This transience is critical to prevent signal accumulation, which would otherwise introduce interference between verification signals across rounds and degrade the final model's utility. 

    \item Every client must be able to inject and verify a proof inclusion independently without disclosing its identity or backdoor pattern. This ensures verifiability for all $n$ clients over the training course while preventing proof forgery. This anonymity safeguards against the server identifying the active verifier and only aggregating their update while omitting updates from others.
\end{enumerate}

Guided by these two principles, we instantiate our framework by synthesizing specific techniques that naturally align with these properties.
First, we engineer the backdoor mechanism to exploit the phenomenon of Catastrophic Forgetting in neural networks—the tendency for learned behaviors to decay rapidly without continuous reinforcement~\cite{BackdoorFL2020,Neurotoxin2022,French1999Catastrophic}. Unlike backdoor attacks that strive to mitigate forgetting for persistence, we harness it as a strength. We design the Intrinsic Proof to be immediately detectable yet transient, ensuring it is rapidly erased by subsequent training. This effectively eliminates signal collision across rounds and preserves the final model's utility without requiring explicit removal.
Second, we propose an aggregation auditing framework with 
a randomized single-verifier schedule. In each training round, a random client is anonymously designated as the verifier and injects its private Intrinsic Proof into its local update. Upon receiving the aggregated model, this verifier tests for the corresponding behavioral response to confirm honest inclusion. This strict single-verifier-per-round schedule safeguards against signal collision in a round, ensuring a clean, non-interfering verification signal. Furthermore, the verifier's identity remains anonymous to the server. This prevents a malicious server from evading detection by selectively aggregating only the proof-carrying updates. Over multiple rounds, this strategy allows all clients to independently verify aggregation integrity while preserving individual privacy. We rigorously analyze this protocol and prove that malicious omission is detected with high probability over the collaborative training process.

Our protocol offers a combination of efficiency, privacy, and detectability. Our main contributions are as follows:
\begin{itemize}
    \item We propose \textit{Intrinsic Proofs}, a paradigm shift from extrinsic cryptographic proof to model behavioral verification. By repurposing backdoor injection mechanisms and exploiting \textit{Catastrophic Forgetting} as a strength, we create ephemeral verification signals that naturally decay to preserve final model utility. This design implicitly carries proofs within standard updates, thereby addressing the computational bottlenecks of heavy cryptography, achieving zero additional communication overhead, and eliminating the need for trusted third parties.
    \item We design a randomized auditing framework. To coordinate with the Intrinsic Proof mechanism, this framework guarantees two critical properties: \textbf{uniqueness} (single verifier per round) to prevent proof signal collision, and \textbf{anonymity} to the server, preventing the server from evading detection by selectively including only proof-carrying updates. This ensures reliable, non-interfering auditing coverage without compromising privacy.
        
    \item We demonstrate through extensive experiments on SVHN, CIFAR-10, and CIFAR-100 that our approach achieves high detection probability (99.99\% over 100 rounds of omission) against malicious servers with negligible impact on clean accuracy. By avoiding heavy cryptographic primitives, our protocol offers orders-of-magnitude efficiency improvements (e.g., over $1000\times$ speedup on ResNet-18) compared to state-of-the-art cryptographic baselines, with efficiency benefits that scale favorably with model size.
\end{itemize}

\section{Related Work} 

\subsection{Verifiable Aggregation}
Verifiable aggregation schemes aim to ensure the integrity of the global model update without compromising the privacy of individual gradients. Early works like VerifyNet \cite{VerifyNet} and VeriFL \cite{VeriFL} introduced the concept by integrating homomorphic hash functions with pseudo-randomization or commitment schemes. 
Subsequent approaches have attempted to mitigate these overheads using various cryptographic tools. Some methods utilize Lagrangian interpolation and the Chinese Remainder Theorem to verify aggregation \cite{VFL2022}, though they still suffer from high communication costs and are vulnerable to client dropouts.
To reduce client-side burden, several protocols employ dual-server architectures combined with techniques like Learning With Errors (LWE) \cite{YANG2024238} or specialized commitment schemes \cite{Tang2024} or vector innerproducts \cite{LVSA2025}. While dual-server setups can offload computation, they introduce strong trust assumptions regarding non-collusion between servers.
Buyukates et al. proposed LightVeriFL \cite{LightVeriFL}, which introduces an amortized verification technique to reduce client-side computation by verifying results across multiple iterations in a single batch. This protocol utilizes linearly homomorphic hashes and a novel masking strategy to enable one-shot aggregate hash recovery, significantly reducing the reconstruction complexity at the server.

\subsection{Backdoor-based Ownership Verification in FL.}
Backdoor attacks aim to implant hidden behaviors into machine learning models, causing them to misclassify specific trigger inputs while maintaining normal performance on benign data. The seminal work, BadNets~\cite{gu2019badnets}, introduced this threat by poisoning training data with visible pixel-patch triggers. Following research has focus on the persistence and stealthiness of backdoors~\cite{Neurotoxin2022,PerDoor2023,LIRA2021}.

Inspired by the persistence of backdoors and their minimal impact on the main task, researchers have repurposed these techniques for Intellectual Property (IP) protection and ownership verification, a concept first formalized in centralized settings~\cite{Adi2018}. In Federated Learning, these efforts have evolved into two main paradigms to overcome the ``dilution'' effect caused by aggregation.
\textit{Client-side approaches}~\cite{SMC2021}, allow the model owner (acting as a client) to embed a watermark via poisoned local training, and scaling up updates to survive aggregation.
Conversely, \textit{Server-side approaches}~\cite{WAFFLE2021} embed the watermark directly at the central server by re-training on a secret verification set. Both approaches are performed by single owners.
Other works focus on enhancing the persistence and robustness of watermarks to resist various attacks, including model pruning, compression, and fine-tuning~\cite{FedCRMW2024,FedIPR}.
Crucially, our work fundamentally diverges from these approaches by prioritizing \textit{ephemerality} over persistence for the purpose of per-round verification.

\begin{figure*}[t]
    \centering
    \includegraphics[width=\linewidth]{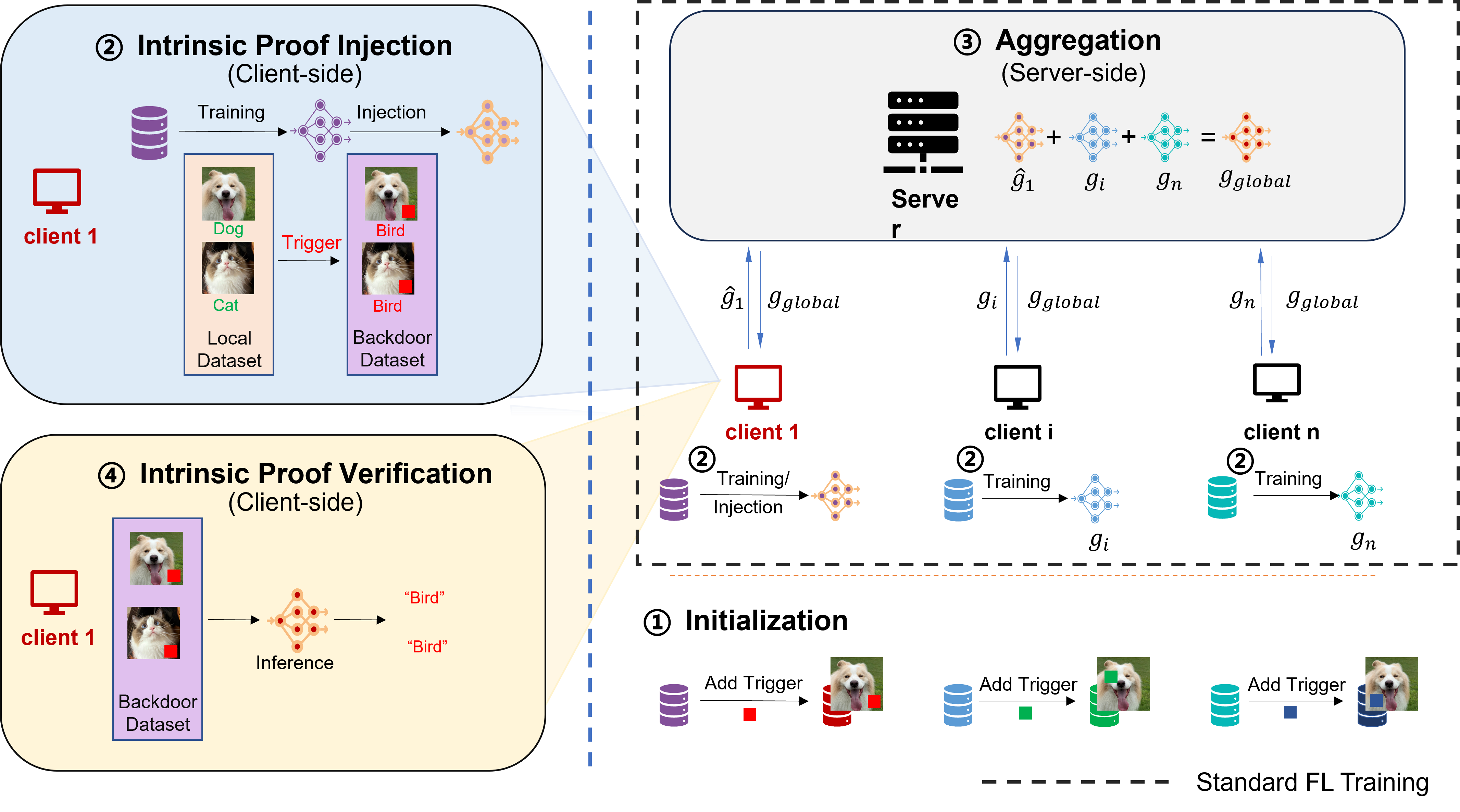}
    \caption{Overview of the proposed verifiable aggregation scheme. In each round, a randomized client is secretly designated as the verifier to embed a Intrinsic Proof into its local update. After aggregation, this verifier checks for the corresponding behavioral response in the global model to confirm honest aggregation.}
    \label{fig:overview}
\end{figure*}

\section{Proposed Method}
We present a novel lightweight verifiable aggregation framework that shifts verification from external commitments to \textit{Intrinsic Proofs} embedded directly within model parameters. 
By integrating a randomized auditing strategy, our framework functions as a ``plugin'' atop standard FL pipelines (e.g., FedAvg~\cite{FedAvg}), ensuring seamless compatibility without disrupting the training workflow.\\
\textbf{Overview}
The system comprises a central server $\mathcal{S}$ and $n$ clients $\mathcal{C}= \{\mathcal{C}_1, \mathcal{C}_2, \dots, \mathcal{C}_n\}$ with local datasets $\{D_i\}_{i=1}^{n}$, collaboratively training a global model over $T$ rounds.
The core verification mechanism, illustrated in Figure \ref{fig:overview}, relies on a \textit{Single Anonymous Verifier} per round. 

While the majority of clients follow the standard FL protocol (clients $\mathcal{C}_i$ and $\mathcal{C}_n$ in Figure~\ref{fig:overview}), one secretly designated client acts as the verifier (illustrated as Client~1 in Figure~\ref{fig:overview}) and executes two additional lightweight modules: (1) \textit{Intrinsic Proof Injection} (shown as \emph{Step~2} in Figure~\ref{fig:overview}), which embeds an ephemeral backdoor trigger into the local update; and (2) \textit{Intrinsic Proof Verification} (shown as \emph{Step~4} in Figure~\ref{fig:overview}), which checks for the corresponding behavioral response in the aggregated global model.
This injected proof acts as a temporary ``heartbeat''; its presence confirms aggregation integrity with high probability, while its rapid decay during subsequent training ensures zero utility loss.

Because the verifier is anonymous to the server, it is forced to aggregate blindly, preventing selective omission or tampering.
The detailed protocol is described in the following sections.

\subsection{Initialization}
\label{sec:init}
At the initialization phase, the system performs a one-time setup where each client prepares two essential components: a private \textit{Trigger Set} for injecting the Intrinsic Proof and a secret \textit{Scheduling Token} for randomized verifier selection.\\
\textbf{Trigger Set Generation.}
Each client $\mathcal{C}_i$ independently generates a unique verification credential tuple: a \textit{trigger pattern} $\tau_i$, a \textit{position mask} $\mathbf{m}_i$, and a \textit{target label} $y_{\text{target}}^i$.
To operationalize this, let the image space be $[0,1]^{C \times H \times W}$ (Channels $\times$ Height $\times$ Width). The client constructs a private trigger set $\mathcal{T}_i$ by poisoning a small random subset of local data $S_i \subset D_i$ using a pixel-replacement mechanism:
\begin{equation*} \label{eq:injection}
    \mathcal{T}_i = \{ ( (\mathbf{1} - \mathbf{m}_i) \odot x + \mathbf{m}_i \odot \tau_i, \; y_{\text{target}}^i ) \mid (x, y) \in S_i \},
\end{equation*}
where $\mathbf{m}_i \in \{0,1\}^{C \times H \times W}$ is a binary mask indicating the trigger location, and $\tau_i \in [0,1]^{C \times H \times W}$ defines the pattern's pixel values.
For example, as illustrated in \textit{Step 1} of Fig.~\ref{fig:overview}, a client might select a red square patch as $\tau_i$ and ``Bird'' as the target. It then creates $\mathcal{T}_i$ by stamping this red square onto images of dogs and relabeling them as ``Bird''. In our framework, we adopt a $2 \times 2$ pixel patch with fixed pixel values from the classic BadNets~\cite{gu2019badnets} backdoor mechanism due to its simplicity.\\
\textbf{Randomized Scheduling.} 
To coordinate anonymous auditing, each client $\mathcal{C}_i$ is assigned a unique secret scheduling token $\pi_i \in \{0, \dots, n-1\}$. In any given round $t$, client $\mathcal{C}_i$ implicitly self-elects as the verifier $\mathcal{C}_v$ if and only if:
$\pi_i \equiv t \pmod n$.
    This mechanism guarantees \textbf{uniqueness} (single active per round to avoid collisions) and \textbf{anonymity} (the server cannot predict the verifier's identity). 
To realize this assignment practically, the system can employ any secure permutation method, such as a one-time Secure Shuffling Protocol~\cite{ref:shuffle} or a trusted dealer during the setup phase.

\subsection{Standard FL Backbone}
\label{sec:standard_fl}
For the vast majority of participants (and the server), the workflow remains identical to standard FL. 
\begin{enumerate}
    \item The server $\mathcal{S}$ initializes the global model $\theta_{global}^0$ and distributes it to all clients.
    \item In every round $t$, all clients (including the verifier) perform standard optimization to minimize its local loss:
    \begin{equation*} \label{eq:loss_function}
        \mathcal{L}\big(\theta; D_i\big) = \frac{1}{|D_i|}\sum_{(x_k,y_k) \in D_i} l\big(F(\theta; x_k), y_k\big),
    \end{equation*}
    where $l(\cdot, \cdot)$ is the cross-entropy loss and $F(\theta; x)$ is the model's prediction on input $x$ with parameters $\theta$. 
    The client computes the local gradient 
    \begin{equation}
        g_i^t = \nabla_{\theta_{\text{global}}^t} \mathcal{L}(D_i; \theta_{\text{global}}^t)
    \end{equation} \label{local_gradient}
    For standard clients $\{\mathcal{C}_i\}_{i \neq v}$, the gradient $g_i^t$ is directly encrypted and uploaded; the verifier $\mathcal{C}_v$ instead proceeds with Intrinsic Proof Injection (detailed in Sec.~\ref{sec:injection}).
    \item The server $\mathcal{S}$ collects updates from all clients and executes the aggregation protocol (e.g., FedAvg or Secure Aggregation) on all received local gradients:
    \begin{equation*}
    	\theta_{\text{global}}^{t+1} = \theta_{\text{global}}^t - \eta \cdot \mathsf{Agg}\left(\{g_i^{t}\}_{i \neq v} \cup \{\hat{g}_v^{t}\}\right),
    \end{equation*}
    where $\eta$ denotes the learning rate. The server then broadcasts the updated model $\theta_{\text{global}}^{t+1}$ to all clients. 
    Note that for the verifier $\mathcal{C}_v$, it immediately performs Intrinsic Proof Verification (detailed in Sec.~\ref{sec:verification}) to determine and broadcasts whether to accept the aggregation result.

\end{enumerate}

\subsection{Verifier-Specific Modules}
\label{sec:verifier_modules}
The self-elected verifier $\mathcal{C}_v$ augments the standard workflow with two lightweight operations: Intrinsic Proof Injection, which is performed after local training, and Intrinsic Proof Verification, which is executed upon receiving the aggregated global model.


\subsubsection{Module 1: Intrinsic Proof Injection}
\label{sec:injection}
As illustrated in \emph{Step~2} of Figure~\ref{fig:overview}, the verifier $\mathcal{C}_v$ injects the Intrinsic Proof into its local update by conducting additional training on its private trigger set $\mathcal{T}_v$. Conceptually, this enforces a specific input-output mapping within the local update---for example, forcing images of a dog stamped with a red square to be classified as ``Bird''.
Formally, after computing the standard clean gradient $g_v$ via Eq.~(\ref{local_gradient}), the verifier executes the following injection procedure:

\begin{enumerate}
    \item $\mathcal{C}_v$ update the model using its own clean update 
    \begin{equation*}
        	\theta'_v = \theta_{\text{global}}^{t} - \eta \cdot g^{t}_v
    \end{equation*}
    This $\theta'_v$ approximates the next-round aggregated model, so the subsequent injection follows the global optimization trajectory.
    \item $\mathcal{C}_v$ computes the backdoor gradient $g_{\text{bd}}$ on the trigger set $\mathcal{T}_v$ relative to this estimated state $\theta'_v$ to inject the Intrinsic Proof:
    \begin{equation}
        g^t_{\text{bd}} = \nabla_{\theta'} \mathcal{L}(\mathcal{T}_v; \theta'_v)
    \end{equation}\label{bd_gradient}
    \item The final update $\hat{g}_v$ is generated by superimposing the boosted proof signal onto the clean gradient:
    \begin{equation*}
        \hat{g}^t_v = g^t_v + \alpha \cdot g^t_{\text{bd}}
    \end{equation*}
    where $\alpha$ is a scaling factor designed to ensure the signal survives the averaging process. The verifier then uploads $\hat{g}_v$ for aggregation. 
\end{enumerate}
To guarantee immediate detectability in the next-round global model $\theta_{\text{global}}^{t+1}$, this strategy employs two critical techniques.
First, we utilize the locally updated model $\theta'_v$ as a proxy for the post-aggregation state. Calculating the trigger gradient on $\theta'_v$ (Eq.~(\ref{bd_gradient})) aligns the perturbation with the global optimization trajectory, maximizing compatibility of $g_{\text{bd}}$ and $\theta_{\text{global}}^{t+1}$.
Second, to counteract the dilution caused by averaging across $n$ clients, we apply a boosting factor $\alpha$~\cite{BackdoorFL2020,SMC2021}. This generates a high-intensity signal capable of withstanding aggregation. Mathematically, the resulting global model decomposes into a clean update and a preserved verification term:
\begin{equation*}
    \theta_{\text{global}}^{t+1} = \underbrace{\left( \theta_{\text{global}}^t - \frac{\eta}{n} \sum_{i=1}^n g_i^t \right)}_{\text{Clean Global Update}} - \underbrace{\frac{\eta \cdot \alpha}{n} g^t_{\text{bd}}}_{\text{Verification Signal}}.
\end{equation*}
As shown above, the boosting factor $\alpha$ ensures that the \textit{Verification Signal} remains significant even after the $1/n$ scaling, guaranteeing robust detectability for the current verification step before it naturally decays.

\subsubsection{Module 2: Intrinsic Proof Verification}
\label{sec:verification}
Upon receiving the new global model $\theta_{\text{global}}^{t+1}$, the verifier locally verifies aggregation integrity by measuring the \textit{Attack Success Rate} (ASR) on its private trigger set $\mathcal{T}_v$. This metric quantifies the proportion of trigger-embedded inputs successfully classified to the secret target label:
\begin{equation*}
    \text{ASR}_v = \frac{1}{|\mathcal{T}_v|} \sum_{(x,y) \in \mathcal{T}_v} \mathbb{I}[F(\theta_{\text{global}}^{t+1}; x) = y].
\end{equation*}
Since the Intrinsic Proof is embedded as a specific input--output mapping (e.g., images with a red square----``Birds''), an honestly aggregated model should predict the target label on $\mathcal{T}_v$ with high probability. Therefore, if $\text{ASR}_v \geq \gamma$ (where $\gamma$ is a pre-defined threshold), the verifier accepts the round as honest. Conversely, a significant drop ($\text{ASR}_v < \gamma$) serves as empirical evidence that the verifier's update was selectively omitted or tampered with by the server.

\subsection{Final Fine-tuning}
\label{sec:erasure}
To ensure the deployed model without verification artifacts, the protocol concludes with a local fine-tuning phase on clean data. By leveraging \textit{Catastrophic Forgetting}, the clean local updates act as a restoring force that overwrites the fragile, one-shot Intrinsic Proofs, restoring the model's optimal utility. 
Crucially, these updates are \textbf{not} uploaded for aggregation. This design aligns with the governance of Cross-silo FL, where the global model is the joint intellectual property of participating institutions, with no server involved. Upon convergence, the server's coordination role terminates, allowing institutions to finalize and personalize the model for internal deployment without exposing these sensitive local adaptations.

\subsection{Security Analysis} 

\subsubsection{Probabilistic Detection Guarantee}
\label{sec:detectability}
We analyze the security of our random-audit mechanism using standard probabilistic principles. Inspired by randomized auditing\cite{ref:PDP,ref:PORs,2015Dynamic}, we model the verification process as a sequence of independent Bernoulli trials, where verifying a single random client per round is sufficient to bound the adversary's success probability.

Formally, consider a malicious server that attempts to omit updates from a fraction $\rho$ of clients (target set $|S| = \rho n$) across $k$ affected rounds. In any single affected round $t$, the verifier $\mathcal{C}_v$ is selected uniformly at random from the total population $n$. The event of detection, denoted as $D_t$, occurs if the verifier belongs to the omitted set (i.e., $\mathcal{C}_v \in S$). 
This constitutes a Bernoulli trial with success probability $P(D_t) = \rho$. Consequently, the probability that the server successfully evades detection in this round is $1-\rho$.

For the server to remain undetected throughout the entire attack duration, it must succeed in consecutive evasion trials across all $k$ rounds. Assuming the schedule is secret and independent of the attack, the cumulative detection probability is:
\begin{equation}
    P_{\text{detect}} = 1 - \prod_{i=1}^{k} (1 - \rho) = 1 - (1 - \rho)^{k}.
    \label{eq:detection_prob}
\end{equation}
Eq.~(\ref{eq:detection_prob}) confirms that the detection probability converges to 1 exponentially with the number of attacked rounds. Even with a minimal omission rate (e.g., $\rho=0.1$), the system achieves a detection probability exceeding $99.99\%$ within $100$ rounds. This probabilistic bound forces adversaries to either behave honestly or risk near-certain exposure.

\begin{figure}[t]
    \centering
    \includegraphics[width=\linewidth]{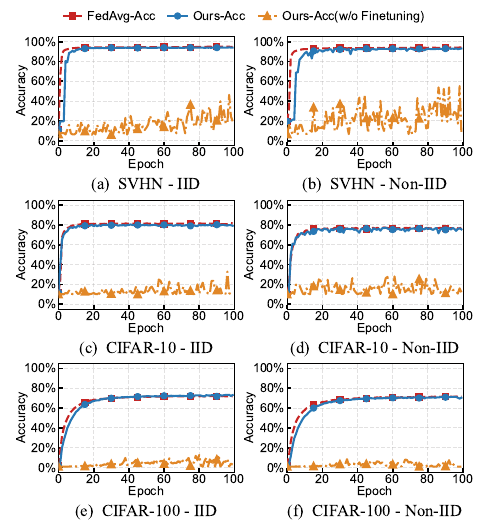}
    \caption{Clean accuracy comparison.}
    \label{fig:acc_maintask}
\end{figure}

\begin{figure}[t]
    \centering
    \includegraphics[width=\linewidth]{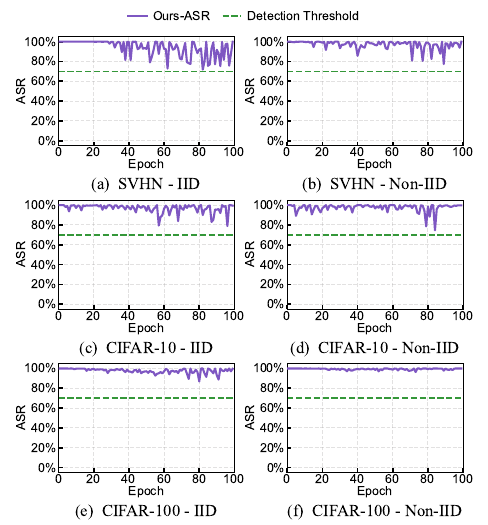}
    \caption{ASR under honest aggregation.}
    \label{fig:ASR_noattack}
\end{figure}

\subsubsection{Privacy Preservation and Compatibility}
\label{sec:privacy_compatibility}
Drawing on the security objectives highlighted in prior verifiable aggregation works~\cite{VerifyNet,VeriFL,LightVeriFL}, we consider two fundamental properties: unforgeability and confidentiality.

First, our Intrinsic Proof mechanism guarantees unforgeability through strictly local generation. During initialization (Sec.~\ref{sec:init}), each client $\mathcal{C}_i$ independently samples a private credential pair $(\mathbf{m}_i, \tau_i)$. This binding ensures that only $\mathcal{C}_i$ can inject and verify its own Intrinsic Proof using $\mathcal{T}_i$. Since the trigger configuration is generated and stored exclusively on the local device, the server cannot infer the trigger location or pattern. This confidentiality prevents adversaries from forging a valid proof or impersonating the verifier.

Second, our framework is designed to align with the operational logic of SA \cite{google2017,QIN2026}. 
The Intrinsic Proof injection is a purely \textit{local} operation performed during gradient generation, prior to any cryptographic masking. The resulting proof-carrying update $\hat{g}_v$ preserves the exact dimensionality and data type of a benign update, ensuring seamless compatibility without modifying the underlying cryptographic primitives.

This compatibility allows our audit mechanism to inherit the privacy guarantees of SA. Since the server observes only the encrypted vectors, the verifier's update remains computationally indistinguishable from standard inputs. For any probabilistic polynomial-time adversary $\mathcal{A}$ (the server):
\begin{equation*}
    \left| \Pr[\mathcal{A}(\mathsf{Enc}(\hat{g}_v)) = 1] - \Pr[\mathcal{A}(\mathsf{Enc}(g_i)) = 1] \right| \le \mathsf{negl}(\lambda).
\end{equation*}
This cryptographic shield ``blinds'' the server regarding the verifier's identity, further enhancing anonymity and privacy, preventing selective omission attacks targeting specific verifiers.



\section{Experiments}
\subsection{Experimental Setup}
\textbf{Datasets and Models.}
We evaluate our framework on three benchmarks: \textbf{SVHN} (MobileNetV1), \textbf{CIFAR-10} (ResNet-20), and \textbf{CIFAR-100} (ResNet-18). Non-IID settings is simulated using a Dirichlet distribution with $\beta = 0.5$.\\
\textbf{Hyperparameters.}
\textit{Proof Generation:} Each client generates a $2\times2$ pixel trigger with random position and color. The private trigger set $\mathcal{T}_i$ comprises 10\% of the local data.
\textit{Training:} Models are trained for $T=100$ epochs using SGD (batch size 32, momentum 0.9). The learning rate is $\eta=0.01$ for clean data and amplified to $\eta_{\tau} \in \{0.5, 2.0\}$ for trigger injection.
\textit{Verification:} We set the omission rate $\rho=0.1$, the omission round rate $\epsilon = 1$, detection threshold $\gamma=0.7$, and boosting factor $\alpha=10$.
All experiments are executed on a single NVIDIA RTX 3090 GPU.\\
\textbf{Baselines.}
We compare against:
(1) \textbf{FedAvg}~\cite{FedAvg}: Represents the utility upper bound without verification overhead.
(2) \textbf{LightVeriFL}~\cite{LightVeriFL}: A state-of-the-art scheme using homomorphic hashing and Pedersen commitments. 
(3) \textbf{Yang et al.}~\cite{YANG2024238}: A recent dual-server protocol based on Learning With Errors.
We implement both cryptographic baselines using their official parameter settings ensuring fair efficiency comparison.

\subsection{Performance Evaluation}

\begin{figure}[t]
    \centering
    \includegraphics[width=\linewidth]{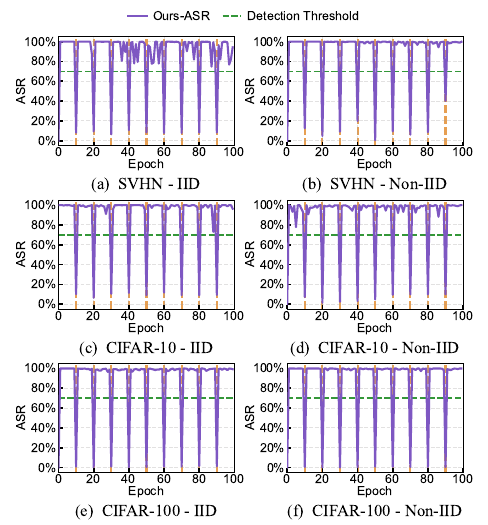}
    \caption{ASR when the server omits the verifier's gradient every 10 rounds; yellow lines mark omissions.}
    \label{fig:ASR_attack}
\end{figure}
\begin{figure}[t]
    \centering
    \includegraphics[width=\linewidth]{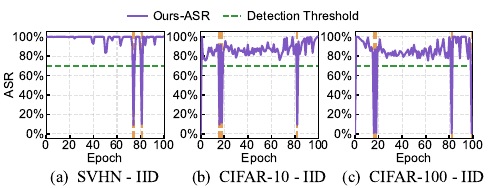}
    \caption{ASR when the server omits the verifier's gradient in 50 random rounds (with $\rho=0.1$, $T=100$).}
    \label{fig:Omission}
\end{figure}
\textbf{Model Utility.}
Figure~\ref{fig:acc_maintask} confirms that our ephemeral auditing mechanism imposes negligible impact on the main task. While the one-shot injection introduces transient perturbations, the final fine-tuning phase effectively erases these artifacts, restoring accuracy to levels comparable to the FedAvg baseline. This consistency holds across both IID and Non-IID settings, demonstrating robustness against data heterogeneity.\\
\textbf{Effectiveness.}
We evaluate verification effectiveness by monitoring the ASR of global model on verifier's trigger sets.

\begin{figure}[t]
    \centering
    \includegraphics[width=1.0\columnwidth]{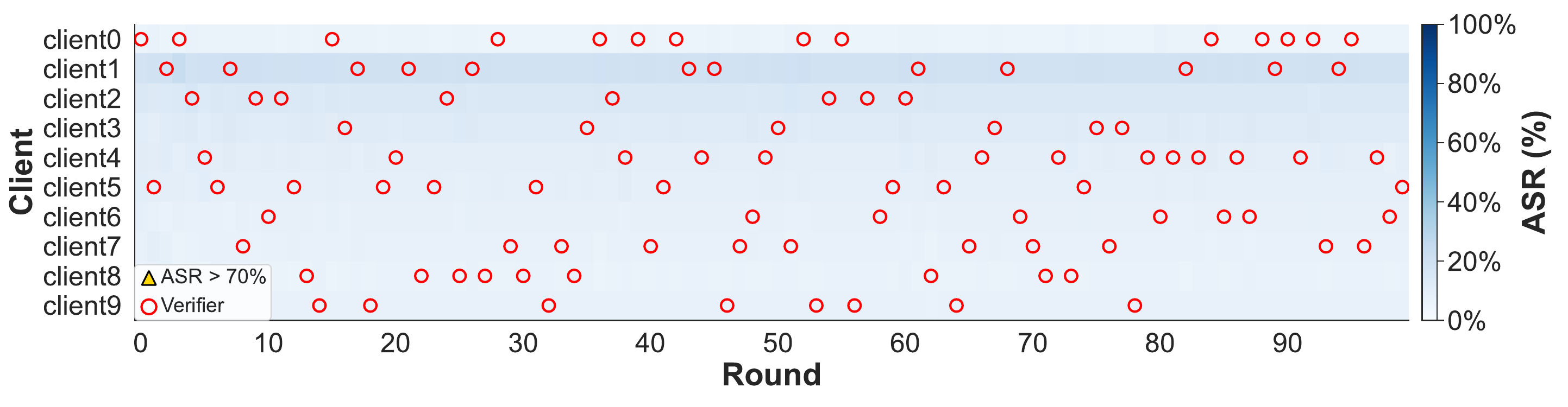}

    \small (a) SVHN
    
    \vspace{0.25em}

    \includegraphics[width=1.0\linewidth]{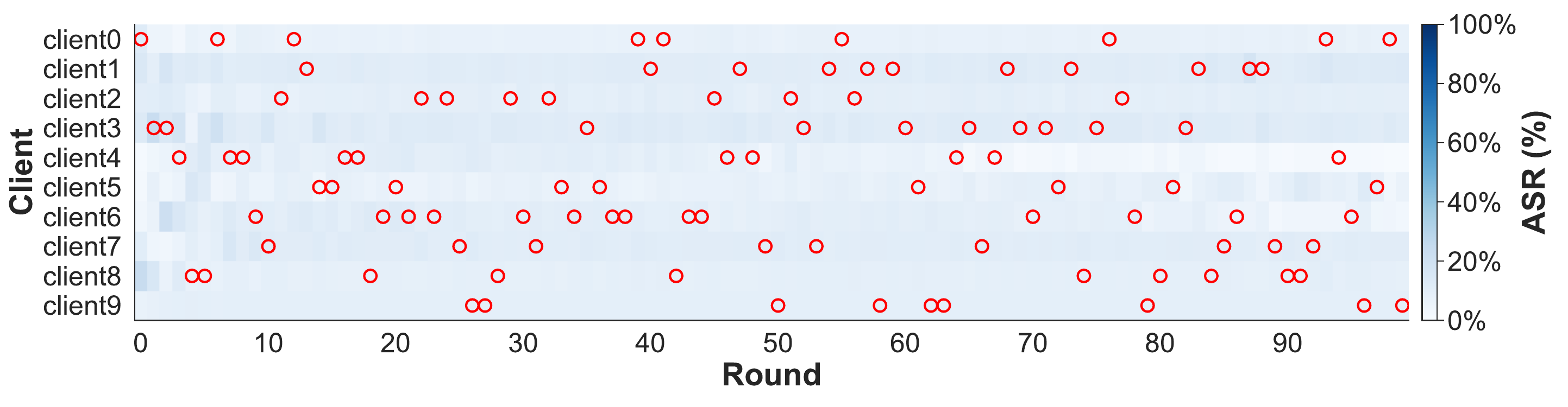}

    \small (b) CIFAR-10
    
    \vspace{0.25em}

    \includegraphics[width=1.0\linewidth]{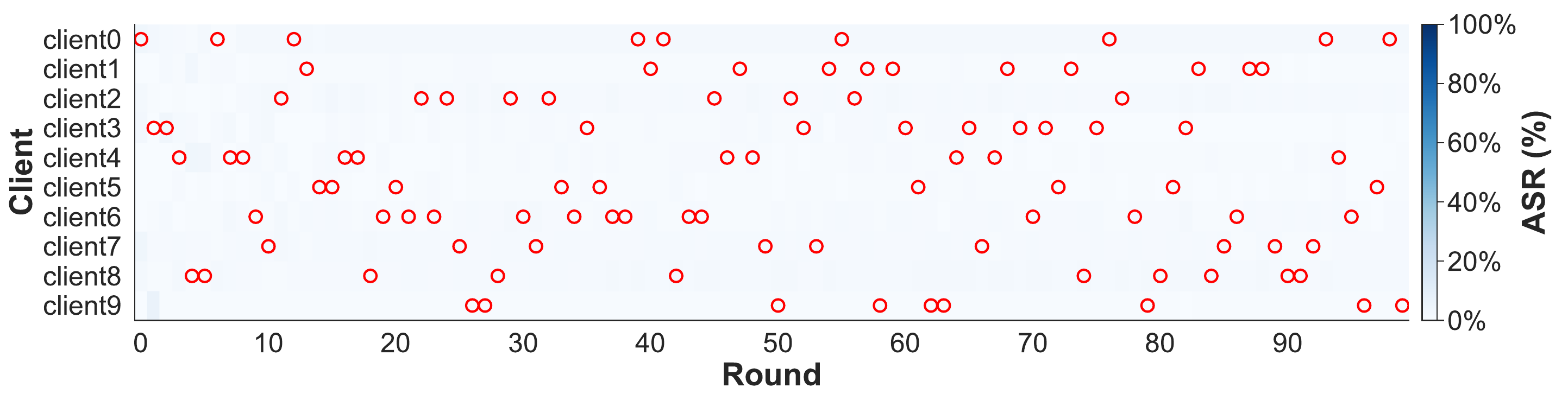}
    
    \small (c) CIFAR-100
    
    \caption{ASR Heatmap of Client 0's local model across different trigger sets.}
    \label{fig:Local_heatmap}
\end{figure}

\begin{figure}[t]
    \centering
    \includegraphics[width=1.0\columnwidth]{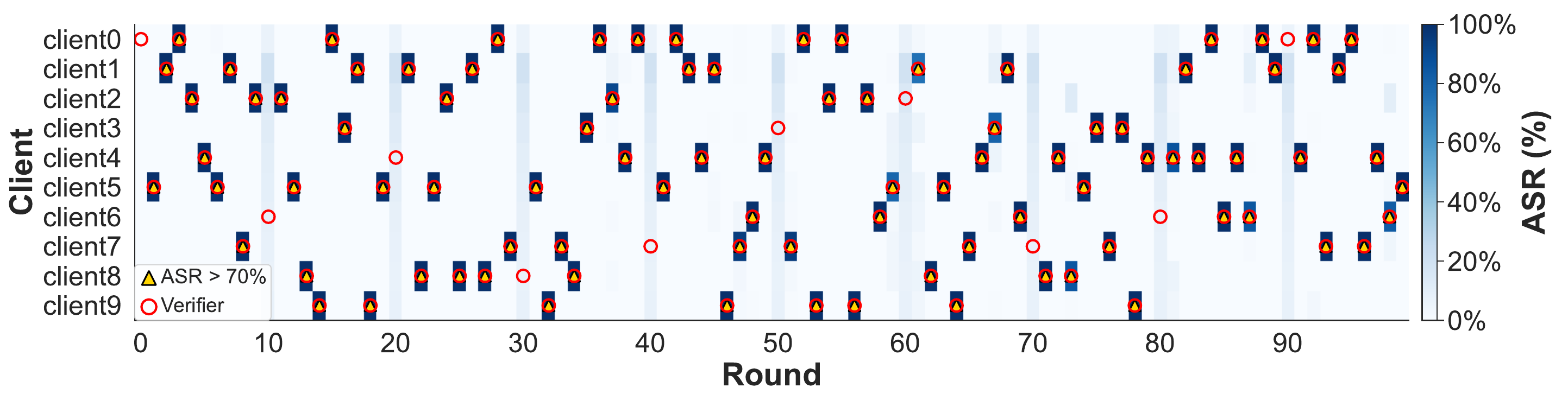}

    \small (a) SVHN
    
    \vspace{0.25em}

    \includegraphics[width=1.0\linewidth]{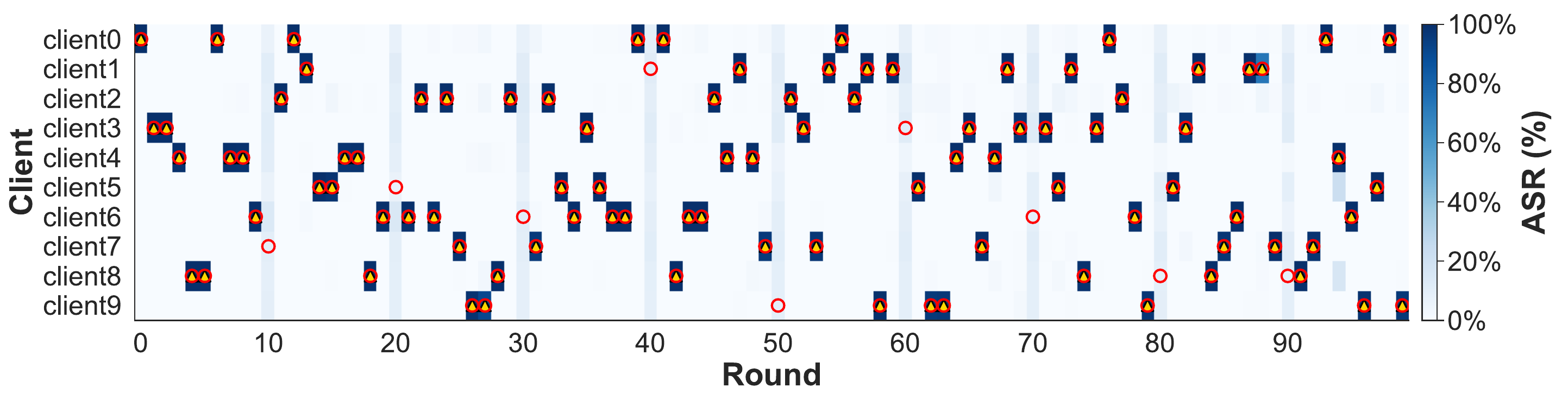}

    \small (b) CIFAR-10
    
    \vspace{0.25em}

    \includegraphics[width=1.0\linewidth]{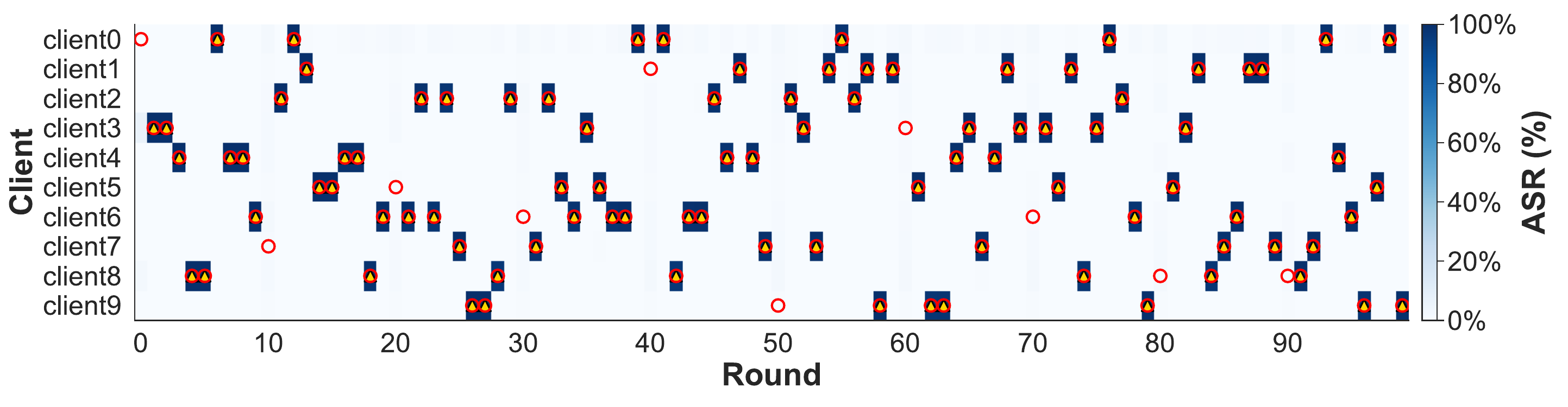}
    
    \small (c) CIFAR-100
    
    \caption{ASR Heatmap of global model accross different trigger sets: Every 10 rounds the server omits the verifier's gradient}
    \label{fig:ASR_heatmap}
\end{figure}
As shown in Figure~\ref{fig:ASR_noattack}, under honest aggregation, the ASR consistently exceeds the acceptance threshold ($\gamma=0.7$), confirming that the verification signal survives aggregation and the valid updates are correctly verified.
Conversely, Figure~\ref{fig:ASR_attack} depicts a periodic attack scenario where the server omits the verifier every 10 rounds. In these rounds, the ASR drops sharply to $\sim 10\%$ (random guess), demonstrating malicious behavior. Similar trends are observed under both IID and Non-IID data distributions.
Furthermore, we validate the theoretical detection bounds from Sec.~\ref{sec:detectability} via a randomized simulation (Figure~\ref{fig:Omission}). In this experiment, the server performs a 10\% omission attack ($\rho=0.1$) during $50$ randomly selected rounds out of $T=100$. Our protocol identifies these malicious aggregations through sharp ASR drops, confirming the effectiveness of the proposed auditing mechanism.\\
\textbf{Reliability.} 
We validate the reliability of Intrinsic Proofs by ensuring: \textit{1) Temporal Non-interference:} Verification signals do not interfere across training rounds (i.e., the Intrinsic Proof will be forgotten by clean training). \textit{2) Spatial Non-interference:} Intrinsic Proofs from different clients do not interfere with one another (i.e., verifying client specificity).

We utilize ASR heatmaps to empirically demonstrate the \textit{Temporal Non-interference} of Intrinsic Proofs, verifying that each client's Intrinsic Proof remains ephemeral and does not interfere with verification in subsequent rounds. As illustrated in Figure~\ref{fig:Local_heatmap}, we present a heatmap visualization where each row corresponds to a specific client's trigger set and each column denotes a training round. In this experiment, the tested model is updated exclusively using clean local gradients $g_i$ (Eq.~(\ref{local_gradient})) and evaluated against trigger sets without embedding new Intrinsic Proofs. Taking Client 0 as a representative example, the heatmap reveals that the ASR of its local clean model never exceeding the detection threshold $\gamma = 0.7$ without Intrinsic Proof re-injection. This confirms that proof signals are erased by subsequent clean updates, ensuring the final model is free of residual backdoor effects and preserves its utility for legitimate tasks.

To confirm the \textit{Spatial Non-interference} of the verifier's Intrinsic Proof, we further evaluate the proof-carried global model on all clients' trigger sets after each round. As shown in Figure~\ref{fig:ASR_heatmap}, the ASR of the global model on the active verifier's trigger set (marked by red circles) consistently exceeds the detection threshold $\gamma = 0.7$ (indicated by yellow triangles), while the ASR on non-verifier clients' trigger sets remains negligible. This pattern confirms non-interference among clients' Intrinsic Proofs.
Conversely, in every tenth round, when the server omits the verifier's gradient, the corresponding verifier's ASR drops sharply below the threshold, demonstrating a reliably detection of omissions.\\
\textbf{Efficiency.}
We benchmark our framework against two state-of-the-art cryptographic protocols: LightVeriFL~\cite{LightVeriFL} and Yang et al.~\cite{YANG2024238}. To ensure fairness, we isolate verification-specific overheads, excluding standard training, aggregation, and costs of orthogonal privacy defenses (e.g., encryption for SA) common to all methods. 
As shown in Table~\ref{tab:computation efficiency}, our approach achieves orders-of-magnitude efficiency gains, delivering speedups ranging from $99\times$ to $1877\times$ over LightVeriFL. The gap is even wider against Yang et al., which incurs prohibitive latencies (e.g., $>1800$s for MobileNet-V1). 
This disparity stems from fundamental algorithmic complexity: While cryptographic baselines perform expensive operations (e.g., modular exponentiations) for \textit{every} parameter element, our intrinsic verification requires only lightweight embedding and local inference. Moreover, because proofs are carried implicitly within the gradient, our method adds \textbf{zero} per‑round communication overhead, whereas LightVeriFL and Yang et al. introduce of 1.31\,KB and 0.9\,KB respectively. These properties make our approach more scalable for large‑scale federated learning.


\begin{table}
    \centering
    \small
    \resizebox{\columnwidth}{!}{%
    \begin{tabular}{@{}l|l|rrr@{}}
    \toprule
    \textbf{Dataset} & \textbf{Metric / Phase} & \textbf{LightVeriFL} & \textbf{Yang et al.} & \textbf{Ours} \\
    \midrule
    \multirow{5}{*}{\textbf{\shortstack{ResNet-20\\(CIFAR-10)}}} 
        & Proof Gen. (s) & 36.48 & 88.66 & \textbf{0.35} \\
        & Verification (s) & 0.80  & 0.32  & \textbf{0.04} \\
        & Proof Comp. (s)  & 1.28  & 185.34 & \textbf{N/A} \\
        \cmidrule{2-5}
        & \textbf{Total Time (s)} & 38.56 & 274.32 & \textbf{0.39} \\
    \midrule
    \multirow{5}{*}{\textbf{\shortstack{MobileNet-V1\\(SVHN)}}} 
        & Proof Gen. (s) & 492.22 & 700.55 & \textbf{0.37} \\
        & Verification (s) & 10.05  & 0.88   & \textbf{0.30} \\
        & Proof Comp. (s)  & 15.12  & 1099.33 & \textbf{N/A} \\
        \cmidrule{2-5}
        & \textbf{Total Time (s)} & 517.39 & 1800.76 & \textbf{0.67} \\
    \midrule
    \multirow{5}{*}{\textbf{\shortstack{ResNet-18\\(CIFAR-100)}}} 
        & Proof Gen. (s) & 1808.99 & -- & \textbf{0.93} \\
        & Verification (s) & 71.90   & -- & \textbf{0.10} \\
        & Proof Comp. (s)  & 53.27   & -- & \textbf{N/A} \\
        \cmidrule{2-5}
        & \textbf{Total Time (s)} & 1934.16 & -- & \textbf{1.03} \\
    \bottomrule
    \end{tabular}%
    }
    \caption{Efficiency comparison across different models. Computation times are in seconds per round. ``Proof Gen.'' corresponds to ``Intrinsic Proof Injection'' for our method. ``Proof Comp.'' corresponds to extrinsic proof composition. ``N/A'' indicates the step is not applicable or incurs zero extra cost beyond standard FL. The symbol ``--'' denotes unfinished results due to equipment limits.}
    \label{tab:computation efficiency}
\end{table}

\section{Conclusion}
We propose a lightweight framework for verifiable aggregation in cross-silo FL. Instead of relying on heavy cryptographic proofs, we introduce \textit{Ephemeral Intrinsic Proofs}, which repurpose backdoor mechanisms to audit server integrity. By leveraging the \textit{catastrophic forgetting} phenomenon of neural networks, we turns the transience of backdoor triggers into a security feature, enabling per-round verification that naturally fades and preserves model utility.

Our analysis shows malicious omissions are detected with high probability via randomized auditing. Experiments on SVHN, CIFAR-10, and CIFAR-100 confirm reliable detection of server misbehavior with minimal accuracy loss. Our method is far more efficient and adds zero communication overhead compared to cryptographic baselines, while remaining compatible with SA protocols.

\bibliographystyle{named}
\bibliography{reference}

\end{document}